\documentclass[10pt,twoside]{hsqcd}
\usepackage{epsf,amsmath}
\usepackage{graphicx}
\usepackage{subfigure}
\begin{document}

\title{Recent results on B physics at Tevatron}

\author{N.~D'Ascenzo \\
{\it for the CDF and D0 collaborations} \\ \\
 LPNHE University Pierre et Marie Curie CNRS/IN2P3, Paris, France\\
Nuclear Research National University, Russia\\
E-mail: ndasc@mail.desy.de }

\maketitle

\begin{abstract}
\noindent The measurement of the CP violation in the b-meson system and of b-meson rare decays provides information about the electroweak symmetry breaking in terms of flavour structure of the CKM matrix and flavour changing neutral currents. The deviation of the experimental observations from the Standard Model predictions allows to constrain new physics. In this paper recent results from Tevatron are reported.
\end{abstract}

\section{CDF and D0 experiments at Tevatron}
Tevatron is a $p\bar{p}$ collider with a center of mass energy of 1.96 TeV. A total luminosity of 8.8 $fb^{-1}$ was delivered to the experiments and 7.3 $fb^{-1}$ was recorded up to July 2010.  

The two experiments at Tevatron, CDF and D0,  are suitable for the analysis of b-mesons properties. 

The CDF detector relies on an excellent tracking system, with momentum resolution $\sigma_{p_{t}}/p_{t}^{2}=0.07\%$[GeV/c]$^{-1}$ and impact parameter resolution $\delta D_{0}=40 \mu$m. A dedicated trigger allows to provide the track displacement information from the silicon detectors with several 10 kHz rates. The electron identification is performed in the calorimeters and muons are identified in the muon system with a coverage up to $\eta=1$. The ionization measurement in the drift chambers combined with time of flight measurement in the TOF detector allow an efficient $\pi/K$ separation used for the identification of kaons from b-quark fragmentation and b-mesons decay.

The key-point of the D0 detector is the excellent coverage of the tracker and of the muon system up to $\eta>2$. The vertex detector was updated in 2006 with an additional silicon layer L0 near the beam pipe (R=1.6 cm). The momentum resolution is $\sigma_{p_{t}}/p_{t}^{2}=0.14\%$[GeV/c]$^{-1}$ and the impact parameter resolution is $\delta D_{0}=13\mu\rm{m}\oplus 50 /p_{t}$. 

\section{New physics in $B_{s}$ oscillation and CP violation}
The $B_{s}\bar{B}_{s}$ system is a two states quantum system with flavour eigenstates (b,$\rm{\bar{s}}$),($\rm{\bar{b}}$,s). 

The flavour eigenstates are not CP eigenstates ($\rm{CP}\psi_{B_{s}^{0}}=-\psi_{\bar{B}_{s}^{0}}$)
 and mixing between the two states can be mediated by the weak interaction through the flavour-changing transition $b\rightarrow s$. 
 
Flavour-changing processes involving quarks are described in the SM by the lagrangian:
\begin{equation}
	\mathcal L=\frac{g}{\sqrt{2}}\left(\hat{u},\hat{c},\hat{t} \right)_{L}\gamma_{\mu}\left(\begin{array}{ccc}V_{ud}&V_{us}&V_{ub}\\V_{cd}&V_{cs}&V_{cb}\\V_{td}&V_{ts}&V_{tb}\\  \end{array} \right) \left( \begin{array}{c}d\\s\\b\end{array}\right)_{L}W_{\mu}^{+}
\end{equation}
The matrix $V_{ij}$ is the Cabibbo-Kobayashi-Maskawa (CKM) matrix~\cite{CKM}. 

The CKM matrix is unitary. The elements of the CKM matrix satisfy the relations $\sum_{i}{V_{ij}V_{ik}^{*}}=\delta_{jk}$ and $\sum_{i}{V_{ij}V_{kj}^{*}}=\delta_{ik}$. The CKM matrix is parameterized by three mixing angles $\theta_{ij}$ and one CP violating phase $\delta$. The CP violation is generated in the SM only by the phase $\delta$ for flavour changing processes involving quarks. CP violating sources are necessary to explain the matter-antimatter asymmetry observed in nature~\cite{Sak}.  Flavour-changing processes in NP models affect the size of the CKM matrix elements and of the CP violating phase.
 
The $B_{s}^{0}\bar{B}_{s}^{0}$ mixing is described by the effective Hamiltonian 
\begin{equation}
	i\frac{\partial}{\partial t} \psi_{B_{s}^{0}}\left(t\right)= \left(\begin{array}{cc}M-i\frac{\Gamma}{2}& M_{12}-i\frac{\Gamma_{12}}{2}\\M_{12}^{\ast}-i\frac{\Gamma_{12}^{\ast}}{2}& M-i\frac{\Gamma}{2}\\ \end{array} \right)\psi_{B_{s}^{0}}\left(t\right)
\label{Heff}
\end{equation}
The off-diagonal terms depend on the elements of the CKM matrix and are sensitive to NP contributions~\cite{Lenz}.  $M_{12}$ receives contribution from massive internal particles. The main SM contribution is the t-quark and NP models predict highly massive particles which can affect sensitively $M_{12}$.  $\Gamma_{12}$ receives contribution from light internal particles, as c- or u- quarks. It is not sensitive to NP which has contribution at higher mass scale. The phase $\phi_{s}=\rm{arg}\left( -\frac{M_{12}}{\Gamma_{12}}\right)$ is also studied. As the SM expectation is $\phi_{s}=\left(0.0041\pm0.0008\right)$,  $\phi_{s}$  is a very sensitive quantity to new physics effects.  

Three experimental observables can be constructed after diagonalization of the Hamiltonian~\ref{Heff} and determination of the two $B_{s,L}^{0}$ and $B_{s,H}^{0}$ mass eigenstates.

The mass difference between the two mass eigenstates $\Delta M_{s}=\left(M_{H}-M_{L}\right)\approx 2 \left|M_{12}\right|$ is sensitive to NP due to the proportionality to $M_{12}$. It was measured by CDF and D0 collaboration using 1~$fb^{-1}$ of data~\cite{DMs}. The CDF and D0 experiment report respectively  $\Delta M_{s}=17.77\pm 0.10\hspace{2mm} \rm{(stat)} \pm 0.07 \hspace{2mm} \rm{(syst)} \rm~{ps}^{-1}$ and  $\Delta M_{s}=18.53\pm 0.93 \hspace{2mm} (stat) \pm 0.30\hspace{2mm}  (syst) \rm~{ps}^{-1}$. The experimental value of $\Delta M_{s}$ and of $\Delta M_{d}$ \cite{Dmb} are combined to determine the CKM matrix element $V_{td}$, which is used to prove the unitarity condition $V_{ud} V^{\ast}_{ub} + V_{cd} V^{\ast}_{cb} + V_{td} V^{\ast}_{tb} = 0$. The experimental result was found in agreement with the SM expectation \cite{DmbDms}.

The width difference between the two mass eigenstates $\Delta \Gamma_{s}=\left(\Gamma_{L}-\Gamma_{H}\right)\approx 2 \left|\Gamma_{12} \right|\rm{cos}\phi_{s}$ is very sensitive to NP due to the correlation with $\phi_{s}$. The golden mode to study this observable is the measurement of the CP violation in the interference between mixing and decay in $B_{s}^{0}\rightarrow J/\psi \phi$. In this decay the relation $V_{us} V^{\ast}_{ub} + V_{cs} V^{\ast}_{cb} + V_{ts} V^{\ast}_{tb} = 0$ is studied from which the angle $\beta_{s}$ is defined as $2\beta_{s}=2arg\left[-\frac{\left(V_{ts}V_{tb}^{\ast}\right)^{2}}{\left(V_{cs}V_{cb}^{\ast}\right)^{2}} \right]$. The SM expectation is $\beta_{s}=0.04\pm 0.01$ rad. It is hence very sensitive to effects of NP\footnote{A relation exists between the angle $\beta_{s}$ and the parameter $\phi_{s}$. NP alters the phase $2\beta_{s}$ to $\phi^{\Delta}_{s}-2\beta_{s}$ and the phase $\phi_{s}$ to $\phi_{s}+\phi^{\Delta}_{s}$. As both $\phi_{s}$ and $2\beta_{s}$ are expected to be very small in the SM, the effects of NP are dominant and of the same size $\phi^{\Delta}_{s}$ for both parameters.} (section 2.1). 

The flavour specific or semi-leptonic CP asymmetries are also sensitive probe of NP. In case $\bar{B}_{s}^{0}\rightarrow f$ and  $B_{s}^{0}\rightarrow \bar{f}$ are forbidden the asymmetry is defined as $a_{sl}^{s}=\frac{\Gamma\left(\bar{B}_{s}\rightarrow f \right)-\Gamma\left(B_{s}\rightarrow \bar{f} \right)}{\Gamma\left(\bar{B}_{s}\rightarrow f \right)+\Gamma\left(B_{s}\rightarrow \bar{f} \right)}=\frac{\Delta \Gamma_{s}}{\Delta M_{s}}\rm{tan}\phi_{s}$. The same relation yields for $B_{d}^{0}$ mesons: $a_{sl}^{d}\equiv \frac{\Gamma\left(\bar{B}_{d}\rightarrow f \right)-\Gamma\left(B_{d}\rightarrow \bar{f} \right)}{\Gamma\left(\bar{B}_{d}\rightarrow f \right)+\Gamma\left(B_{d}\rightarrow \bar{f} \right)}=\frac{\Delta \Gamma_{s}}{\Delta M_{d}}\rm{tan}\phi_{d}$. The relation to $\phi_{q}$ makes this observables sensitive to NP. These observables are studied in the measurement of the di-muon charge asymmetry\footnote{ $N_{b}^{++}$ and $N_{b}^{--}$ are the number of events in which two selected muons in the final state have the same charge}  $A^{b}_{sl}\equiv\frac{N_{b}^{++}-N_{b}^{--}}{N_{b}^{++}+N_{b}^{--}}$. $A^{b}_{sl}$ receives contribution from both $B_{s}$ and $B_{d}$ mesons: $A_{sl}^{b}=\left(0.506\pm0.043 \right)a_{sl}^{d}+\left(0.494\pm0.043 \right)a_{sl}^{s} $.  The SM expectation is $A_{sl}^{b}=\left(-2.3^{+0.5}_{-0.6}\right)\times 10^{-4}$. NP contributions to the Feynman box diagrams responsible for $B_{q}^{0}$ mesons mixing affect significantly the value of $A_{sl}^{b}$ (section 2.2).

\subsection{CP violation in $B_{s}\rightarrow J/\psi \left(\mu\mu \right)\phi \left(KK \right)$}
The analysis is performed on the decay $B_{s}^{0}\rightarrow J/\psi \phi$ with the subsequent two body decays $\phi\rightarrow KK$ and $J/\psi \rightarrow \mu\mu$. 

The final state of the decay $B_{s}\rightarrow J/\psi \phi$ is composed of a mixture of CP eigenstates. As the $J/\psi$ and $\phi$ are spin-1 particles, the allowed CP states are defined by the relation $CP \psi_{J/\psi \phi} = \eta_{J\/psi}\eta_{\phi}\left(-1 \right)^{L}$, with $L=0,1,2$ and $\eta$ intrinsic CP of the two particles. Two CP-even (L=0,2) and one CP-odd (L=1) states are formed. The total decay amplitude is described in the transversity base~\cite{trans} and is decomposed in the polarization amplitudes $A_{0}$, $A_{\parallel}$ and $A_{\bot}$.

Five parameters defined in QCD are used to describe the polarization amplitudes~\cite{dec}. They are  the strong phase and the amplitude of $A_{\parallel}$, the strong phase and the amplitude of $A_{\bot}$  and the real-valued amplitude $A_{0}$. 

The time evolution of each polarization amplitude is calculated from the Schr\"{o}dinger equation \ref{Heff}. The angular distribution of the particles in the final states is calculated through the decomposition of the $J/\psi\rightarrow \mu\mu$ and $\phi\rightarrow KK$ decay amplitudes in partial amplitudes \cite{ang}. 

The analysis consists of a likelihood fit to the angular and time distributions in order to extract the value of $\Delta \Gamma$ and $\beta_{s}$. This report refers to the recent CDF update with a data sample of $5.2$~fb$^{-1}$ \cite{bsjpsiphi}. The analysis relies on:
\begin{itemize}
	\item The event reconstruction and event selection which yield $6504\pm85$ $B_{s}^{0}\rightarrow J/\psi \phi$ signal events.
	\item Two b-tagging methods. The opposite side tagging combines the information of jet charge and lepton identification to reconstruct the flavour of the b quark at the moment of production. It was calibrated using $B^{+}\rightarrow \Psi K^{+}$ decay and has an efficiency of 94.2 $\%$ and dilution $11.5 \%$. The same side tagging identifies a $K^{\pm}$ from fragmentation of the b-quark. It was calibrated with a simultaneous measurement of $\Delta M_{s}$ and of a dilution scale factor $A$. It has a total tagging power of $3.2 \%$.
	\item Angular efficiency function used to take into account the detector angular acceptance.
\end{itemize}

The resulting value of $\Delta\Gamma_{s}$ and $\beta_{s}$ (Fig. \ref{CPV}a) is found in agreement with the SM prediction at 0.8 $\sigma$. The possibility of the extension of this analysis including data collected with a two displaced tracks trigger is being studied. 

\subsection{CP violation in di-muon charge asymmetry measurement}
\begin{figure}[t]
	\centering
	\subfigure[]{	\includegraphics[width=0.4\textwidth]{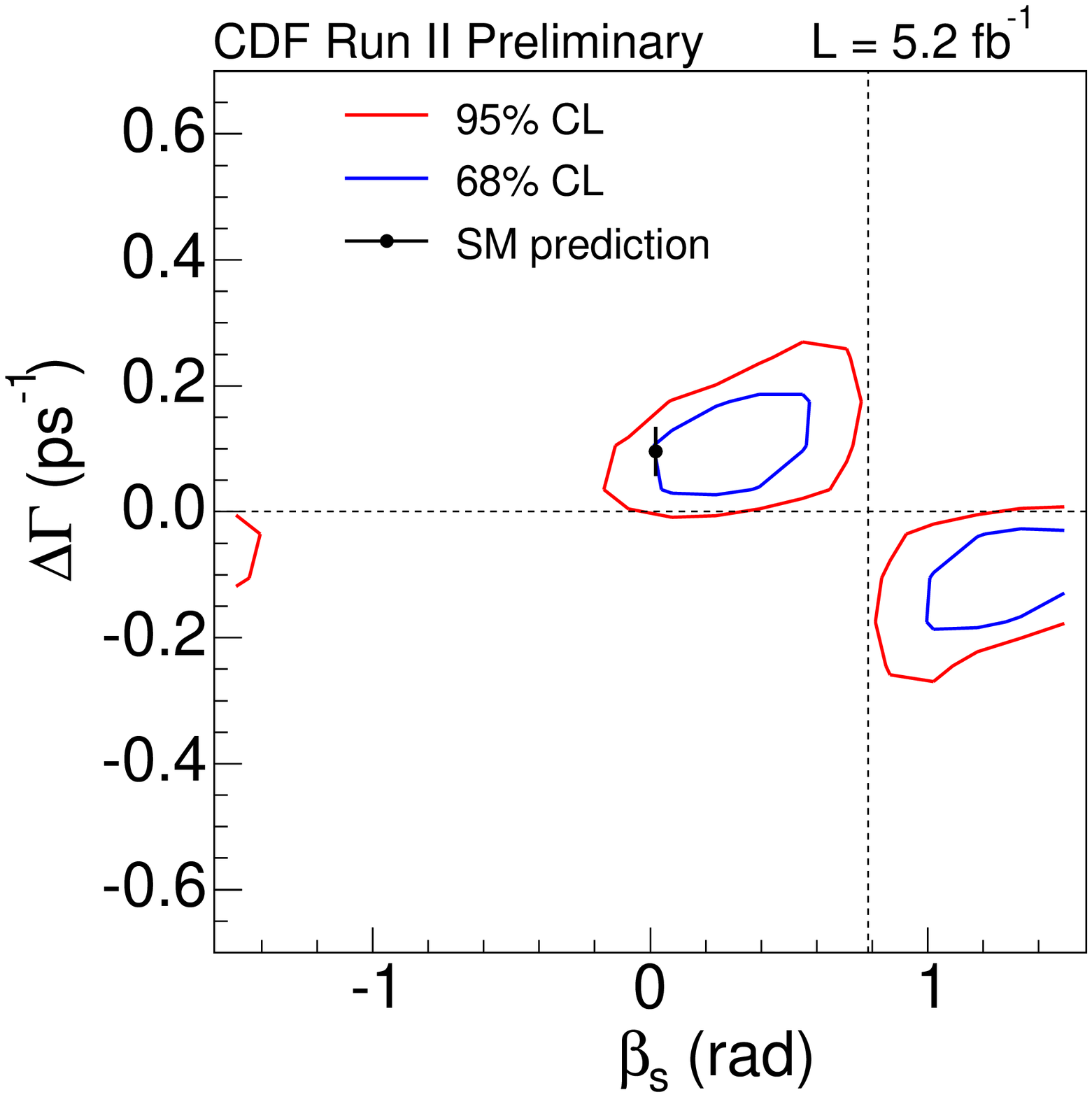}}
	\subfigure[]{	\includegraphics[width=0.4\textwidth]{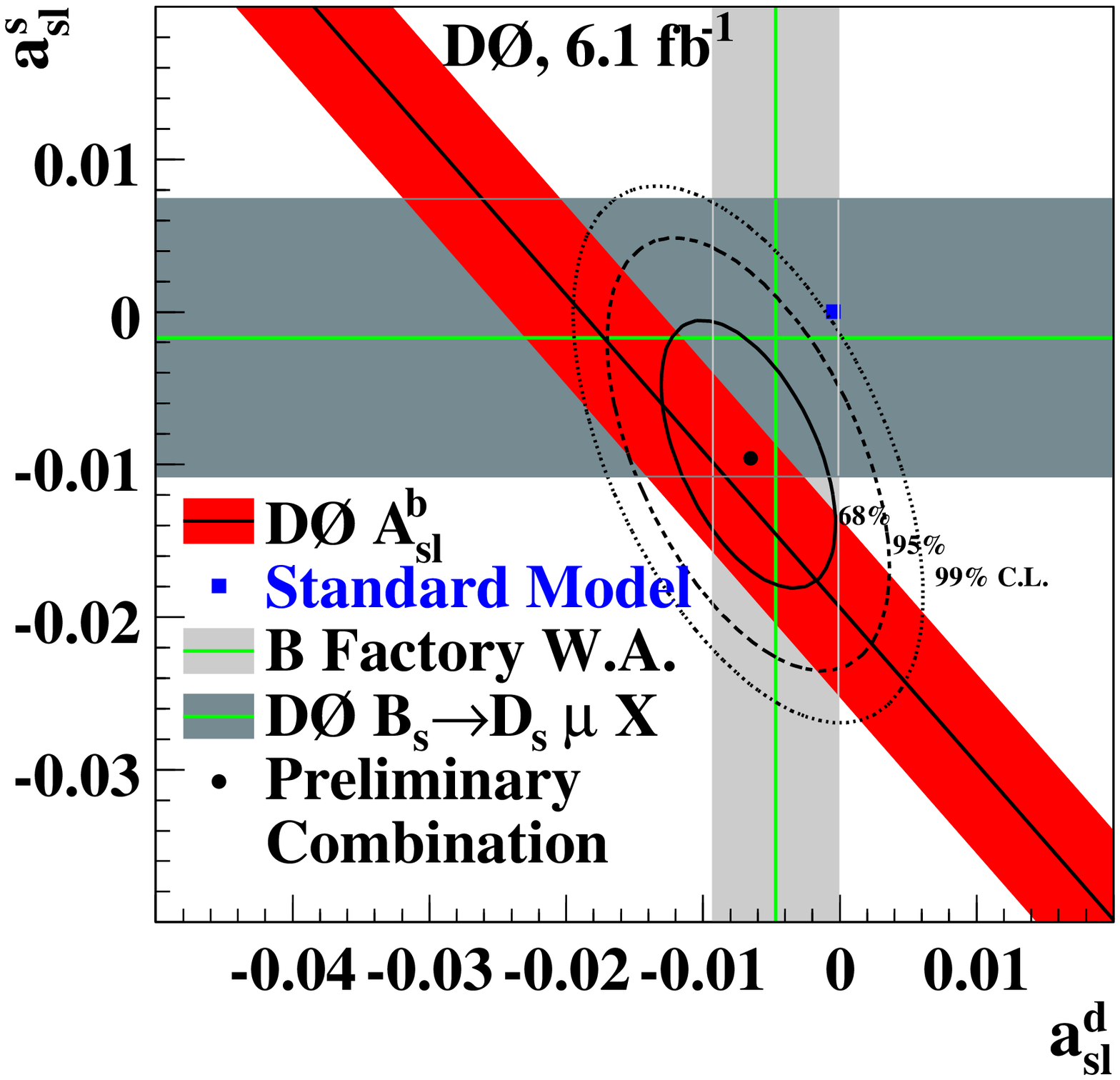}}
	\caption{\label{CPV}Measurement of CP violation in B meson mixing. In (a) the  68$\%$ and 95$\%$ CL contours in the $\beta_{s}--\Delta\Gamma_{s}$ plane are shown, from the CDF analysis of $B_{s}^{0}\rightarrow J/\psi \phi$ decay. A deviation from the standard model expectation (black point) of 0.8$\sigma$ is measured. In (b) the comparison of the  like-sign dimuon charge asymmetry $A_{sl}^{b}$ measured by D0 and the SM is shown. Also shown are the existing measurements of $a_{sl}^{d}$ and $a_{sl}^{s}$. $A_{sl}^{b}$ is found with a deviation of $3.2\sigma$ from the SM.}
\end{figure}
The analysis consists of the reconstruction of events with two same charge muons in the final state. 

In this paper the results are reported from the D0 experiment obtained with a total data set of $6.1$~fb$^{-1}$~\cite{bsasym}.

The analysis is performed on two data samples. The inclusive  muon sample is composed of events with at least one single muon candidate passing the muon selection and at least one single muon trigger. The like-sign di-muon sample consists of all events with at least two muons candidates of the same charge that pass the dimuon selection and at least one dimuon trigger. This analysis relies on:
\begin{itemize}
	\item Reversing of the polarity of the toroidal ad solenoidal magnet every two weeks, in order to cancel first order effects related with the instrumental asymmetry.
	\item Measurement of background contribution to the asymmetry. Muons are considered from the decay of charged kaons and pions and  punch-through kaons, pions and protons. The contribution of these backgrounds to the asymmetry $a$ and $A$ are extracted directly from data.
\end{itemize}

The quantities $a\equiv\frac{n^{+}-n^{-}}{n^{+}+n^{-}}$, where $n^{\pm}$ is the number of positive and negative identified muons,  and $A\equiv\frac{N^{++}-N^{--}}{N^{++}+N^{--}}$ are extracted respectively from the inclusive and like-sign sample. The value of $A_{sl}^{b}$ is extracted independently from $A$ and $a$ according to the relations $a=k\times A_{sl}^{b}+a_{bkg}$ and $A=K\times A_{sl}^{b}+A_{bkg}$ where $a_{bkg}$ and $A_{bkg}$ are the background contributions extracted from data and $k$ and $K$ are scale factors estimated in MC. It was found that $a_{bg}\sim A_{bg}$ so that a subtraction of the two relations reduces the systematic error on $A_{sl}^{b}$. 

The measured like-sign dimuon asymmetry is $A_{sl}^{b}=-0.00957\pm 0.00251\rm{(stat)}\pm0.00146\rm{(syst)}$ (Fig. \ref{CPV}b).  This asymmetry is in disagreement with the prediction of the SM by 3.2$\sigma$ deviation. This is the first evidence of anomalous CP-violation in the mixing of neutral B-mesons.

\section{New physics in rare B meson decays}
\begin{figure}[t]
	\centering
	\subfigure[]{	\includegraphics[width=0.3\textwidth]{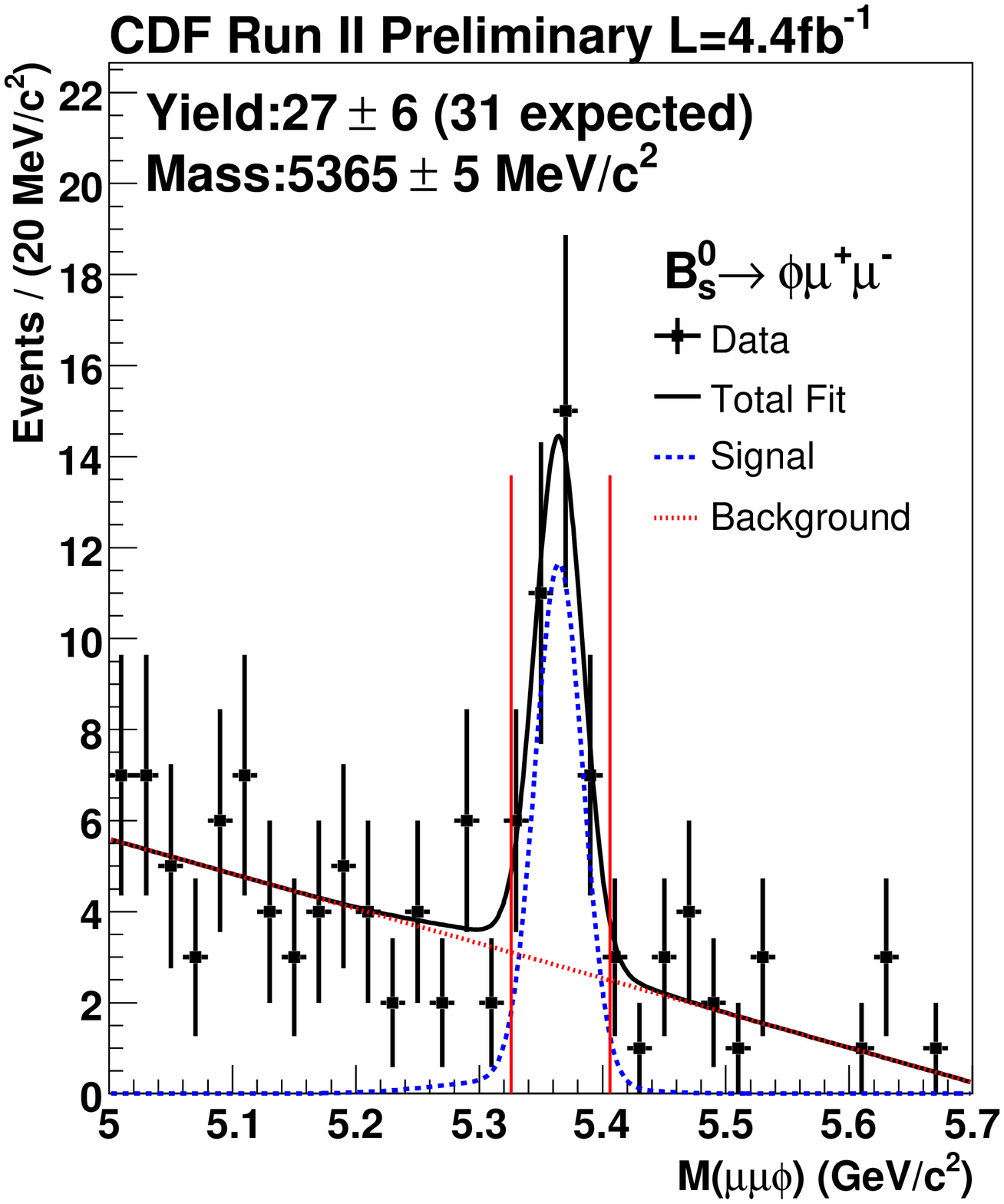}}
	\subfigure[]{	\includegraphics[width=0.48\textwidth]{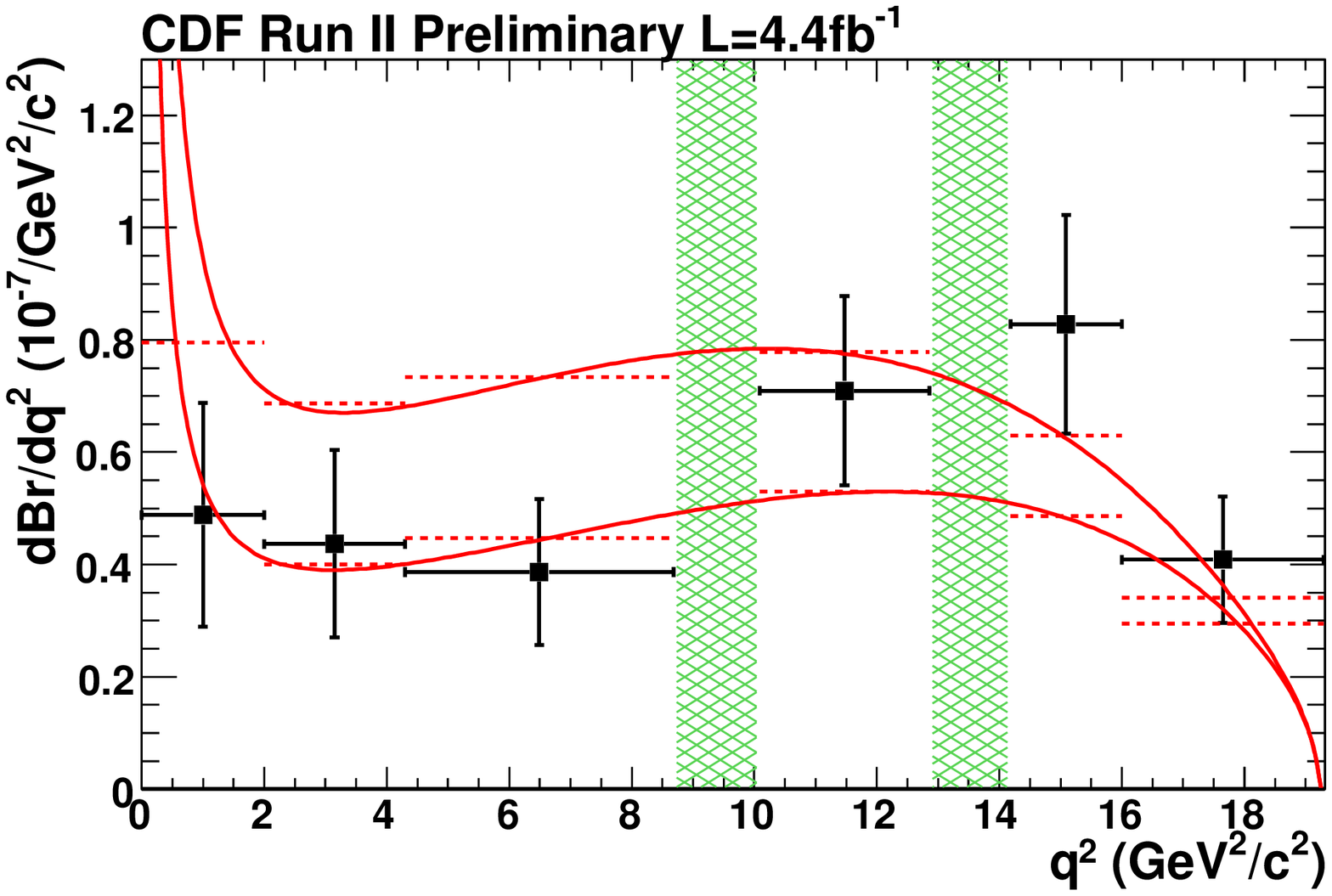}}
	\caption{\label{rare} Rare decays. In (a) the invariant mass is shown of the $B_{s}$ in the first observation of $B_{s}\rightarrow \phi\mu\mu$. In (b) the measurement is shown of the differential branching ratio of $B^{0}\rightarrow K^{\ast 0}\mu^{+}\mu^{-}$, where the solid lines are the SM expectation, the black dots are the data points.}
\end{figure}
\subsection{$B_{s}^{0}\rightarrow \mu^{+}\mu^{-}$}
The decay $B_{s}\rightarrow \mu\mu$ is a Flavour Changing Neutral Current (FCNC) process forbidden at tree level. It occurs through GIM suppressed box or penguin diagrams. The expected branching ratio of this decay in the SM is $BR\left(B_{s}\rightarrow \mu\mu\right)=\left(3.42\pm0.54 \right)\times 10^{-9}$. Decay amplitudes can be enhanced of few orders of magnitude in extensions of the SM. This rare decay is hence very sensitive to effects of new physics.

The analysis is performed on a data sample composed of two identified muons, satisfying the di-muon trigger and the di-muon selection criteria. The number of events passing the selection is normalized to the yield of the well studied normalization channel $B^{+}\rightarrow J/\psi K^{+}$. 

The upper limit on $BR\left(B_{s}\rightarrow \mu\mu\right)$ measured by the two experiments is $4.3\times 10^{-8}$ (CDF \cite{CDFbsmumu}) and $5.1\times 10^{-8}$ (D0 \cite{D0bsmumu}) at $95\%$ C.L. obtained analysing respectively a data sample of $3.2$~fb$^{-1}$ (CDF) and $6.1$~fb$^{-1}$ (D0). The result is in agreement with the SM. 

\subsection{$B_{q}^{0}\rightarrow h \mu\mu$}
The $B\rightarrow h \mu\mu$ decay is a FCNC process, mediated by the quark transition $b\rightarrow s l l$. It is described by three dominant Wilson coefficients $C_{7}^{eff},C_{9}^{eff},C_{10}^{eff}$, where $C_{7}^{eff}$ gets contribution from the photon penguins, $C_{9}^{eff}$ comes from the vector and $C_{10}^{eff}$ form the axial component of the weak diagrams \cite{Wilson}. The BR and the front-backward asymmetry $A_{FB}$ depend on the Wilson coefficients, which indicate if the underlying dynamics is described by SM or NP physics like supersymmetry, Technicolor or fourth generation.

In the CDF analysis performed on a data sample of about $4.4$~fb$^{-1}$ \cite{bkmumu} the decay modes $B^{+}\rightarrow K^{+} \mu^{+}\mu^{-}$, $B^{0}\rightarrow K^{0} \mu^{+}\mu^{-}$ are searched and fully reconstructed. The branching ratio is calculated respect to the normalization channel $B\rightarrow J/\psi h$. Both BR and $A_{FB}$ for the $B^{+}\rightarrow K^{+} \mu^{+}\mu^{-}$, $B^{0}\rightarrow K^{0} \mu^{+}\mu^{-}$ are measured in agreement with the SM and with the current best results at BaBar and Belle (Fig.~\ref{rare}b).

A first observation  with $6.3\sigma$ evidence of the decay mode $B_{s}\rightarrow \phi \mu \mu$ is performed in the CDF experiment on a data sample of about $4.4$~fb$^{-1}$ (Fig.~\ref{rare}a). The measured BR is $BR\left(B_{s}\rightarrow \phi \mu \mu \right)=\left(1.44\pm0.33\rm{(stat)}\pm0.03\rm{(syst)} \right)\times 10^{-6}$ \cite{bkmumu}.
 
\section{Conclusions}
The high luminosity reached at Tevatron and the interplay between CDF and D0 experiments allow high accuracy in the measurement of CP violation in the b-mesons mixing and of rare b-meson decays. Both measurements are important for the constrain of new physics processes and for the understanding of the Standard Model.

\end{document}